\documentclass[prl,aps,twocolumn,superscriptaddress,showpacs,amsmath,amssymb,floatfix]{revtex4}
                                                                                \usepackage{graphicx}

\newcommand{\be}{\begin{equation}}
\newcommand{\ee}{\end{equation}}

\begin{document}

\title{Equilibrium and dynamics of a trapped superfluid Fermi gas with unequal masses}
\author{G. Orso}
\affiliation{Laboratoire Physique Th\'eorique et Mod\`eles Statistiques,
Universit\'e Paris Sud, Bat. 100, 91405 Orsay Cedex, France}
\author{L.P. Pitaevskii}
\affiliation{Dipartimento di Fisica, Universit\`a di Trento and
BEC-INFM, 1-38050 Povo, Italy}
\affiliation{Kapitza Institute for Physical Problems, 117334 Moscow, Russia}
\author{S. Stringari}
\affiliation{Dipartimento di Fisica, Universit\`a di Trento and
BEC-INFM, 1-38050 Povo, Italy}
\date{\today}

\begin{abstract}
Interacting  Fermi gases with  equal populations but unequal masses  are investigated at zero temperature using local density approximation and the hydrodynamic theory of superfluids in the presence of harmonic trapping.  
 We derive the conditions of energetic stability of the superfluid  configuration with respect
to phase separation and the frequencies of the collective oscillations in terms of the mass ratio and the trapping frequencies of the two components. We discuss the behavior of the gas after the trapping potential of a single component is switched off and show that, near a Feshbach resonance, the released component can still remain trapped due to many-body interaction effects. Explicit predictions are presented for a mixture of $^6$Li and $^{40}$K with resonant interaction.

\end{abstract}
\maketitle

The superfluid behavior of dilute interacting Fermi gases at very low temperature is now rather well understood both from the experimental and theoretical point of view \cite{rmp}. In particular the attractive nature of the interaction between the two different spin components  of the gas is known to play a crucial role along the whole BCS-BEC crossover. This includes the case of small and negative values of the scattering length, where the ordinary BCS regime of superfluidity holds,  the BEC regime characterized by the formation of molecules in the Bose-Einstein condensed state and the unitary regime where the scattering length takes a divergent value and the attraction results in  peculiar many-body effects. 
 
A more recent and intriguing direction is the search for superfluidity in mixtures of Fermi gases  
belonging to different species, and hence having different masses. Experimentally, 
the most promising candidates are ultracold mixtures of  $^{40}$K and  $^6$Li, where the mass ratio is $6.7$, 
near a heteronuclear s-wave Feshbach resonances \cite{grimm}.
Equilibrium configurations of the uniform superfluid phase where the atom densities of the two species are equal but the masses are different 
have been theoretically investigated in Ref.\cite{caldas} within BCS mean field theory 
and, more recently,  by quantum Monte Carlo methods \cite{dorotee}.
Other recent works have explored the interplay between different masses and different populations of the two 
components \cite{yip,sademelo}.

In this Letter we investigate the equilibrium and the dynamic properties of a trapped Fermi gas with unequal masses  using local density approximation  and developing the hydrodynamic theory of superfluids at zero temperature.  We assume that the external potentials for the two components, 
hereafter called $\uparrow$ and $\downarrow$, are harmonic and given by 
$V_{\textrm{ho}}^\sigma(\mathbf r)=m_\sigma ( \omega_{x \sigma }^2 x^2+\omega_{y \sigma }^2 y^2+   \omega_{z \sigma }^2 z^2)/2$
where $\sigma=\uparrow,\downarrow$ and $m_\uparrow$ and $m_\downarrow$ are the atomic masses.
Since the two species have different magnetic and optical properties, the trapping frequencies 
$\omega_{i \sigma }$   can be tuned separately.
We consider a mixture of two different fermionic species with equal
populations $N_\uparrow=N_\downarrow=N/2$,  corresponding to 
the most favorable condition for Cooper pairing. We assume that the gas is superfluid at zero temperature and we explore configurations where the densities of the two components are equal and move in phase: $n_\uparrow=n_\downarrow \equiv n/2$, ${\bf v}_\uparrow={\bf v}_\downarrow\equiv {\bf v}$. We will not consider here exotic polarized phases, like the FFLO phase, whose relevance  for trapped Fermi gases is still unclear at present.

At equilibrium, where $\mathbf v=0$,  the atomic density profile $n_0(\mathbf r)$ of the gas 
is given by the local density (also called Thomas-Fermi) approximation for the chemical potential
\begin{equation}
\mu_0 =\mu(n_0(\mathbf r))+\tilde V_\textrm{ho}(\mathbf r),\label{lda}
\end{equation}
where $\mu_0$ is fixed by the normalization condition  $\int n_0(\mathbf r)d\mathbf r=N$ 
and  $\mu (n)=\partial e/\partial n$ is the chemical potential of uniform matter, $e(n)$ being the energy   per unit volume of the homogeneous phase. In Eq.(\ref{lda}) we have introduced the {\sl effective} trapping potential 
\begin{equation}
 \tilde V_{ho}(\mathbf r)=\frac{1}{2}(V_\textrm{ho}^\uparrow +V_\textrm{ho}^\downarrow)\frac{m}{2}(\tilde \omega_{x \sigma }^2 x^2+\tilde \omega_{y \sigma }^2 y^2+  \tilde \omega_{z \sigma }^2 z^2) \label{vtilde},
\end{equation}
given by the average of the two  potentials  and 
\begin{equation}\label{weff}
 \tilde \omega_i^2 = \frac{m_\uparrow \omega_{i \uparrow}^2+m_\downarrow \omega_{i \downarrow}^2 }{m_\uparrow+m_\downarrow}
\end{equation} 
with $m=(m_\uparrow+m_\downarrow)/2$.
In the superfluid phase the densities of the two components are  equal, even if the trapping potentials, in the absence of interactions, would give rise to different equilibrium profiles at zero temperature, i.e. even if the 
oscillator lengths  $\hbar/(m_\sigma \omega_\sigma)^{1/2}$
of the two components do not coincide. If the oscillator lengths are equal, the effective frequencies (\ref{weff}) simply reduce to the geometrical averages $\tilde \omega_i=\sqrt{\omega_{i \uparrow}\omega_{i \downarrow}}$.

Let us discuss the behavior of the equation of state along the BCS-BCS crossover and the corresponding shape of the density profiles. At unitarity, where the scattering length $a$ diverges, the  equation of state takes the universal form \cite{rmp}
$\mu=(3 \pi^2)^{2/3} \hbar^2 (1+\beta) n^{2/3}/4 m_r$, where $m_r=m_\uparrow
m_\downarrow/(m_\uparrow+m_\downarrow)$ is the reduced mass and $\beta$ is a dimensionless parameter depending on the mass ratio $m_\uparrow/m_\downarrow$ and accounting for the interacting effects in this strongly interacting regime. 
By inserting  $\mu=\alpha n^{2/3}$ in Eq.(\ref{lda}), with $\alpha=(3 \pi^2)^{2/3} (1+\beta)/4 m_r$, we find that the density distribution 
of the resonant gas takes the usual form $n(\mathbf r)=(\mu_0-\tilde V_\textrm{ho}(\mathbf r))^{3/2}/\alpha^{3/2}$, 
the Thomas-Fermi radii $R_i$, where the density vanishes, being given by
\begin{equation}\label{RTF}
R_i=\tilde a_\textrm{ho}(24 N)^{1/6}(1+\beta)^{1/4}\frac{\tilde \omega_\textrm{ho}}{\tilde \omega_i},
\end{equation}
where $\tilde \omega_\textrm{ho}=(\tilde \omega_x \tilde \omega_y \tilde \omega_z)^{1/3}$ is the geometrical average 
of the three effective oscillator frequencies and
$\tilde a_\textrm{ho}^2=\hbar/\tilde \omega_\textrm{ho}(m_\uparrow m_\downarrow)^{1/2}$. The  value of $\beta$ is known for the special case of equal masses 
$m_\uparrow = m_\downarrow$, where $\beta =-0.58$ \cite{qmc1,qmc2}. Preliminary  Monte Carlo calculations suggest that $\beta$ depends very weakly on the mass ratio \cite{stefano}. 
In the deep BCS limit, corresponding to a weakly attractive interaction ($n |a|^3\ll 1$), the equation of state
and the Thomas-Fermi radii are given by the same expressions holding at unitarity by simply setting  $\beta=0$. 
 
In the limit of small and positive scattering length, corresponding to $n a^3 \ll 1$, the gas
instead corresponds to a  BEC of diatomic heteronuclear molecules
of mass $m_\uparrow +m_\downarrow=2m$ and molecular density $n/2$. In this regime the 
equation of state is fixed by the repulsive interaction between molecules
and takes the usual bosonic form $\mu_m=g_m n/2$, where the coupling constant $g_m$ is related to the
molecule-molecule scattering length $a_m$ by   $g_m=2\pi \hbar^2 a_m/m$.
The exact value of the molecular scattering length   has been calculated by Petrov et al.\cite{a_molecule} 
as a function of the atom scattering length $a$ and the mass ratio $m_\downarrow/m_\uparrow$. 
The bosonic chemical potential is related to the fermionic one by 
$\mu_m=-E_b+2 \mu$, where $E_b=-\hbar^2/ 2 m_r a^2$ is the two-body binding energy.
From Eq.(\ref{lda}), the density profile is then given by
$n(\mathbf r)=(\mu_0-\tilde V_\textrm{ho}(\mathbf r))2m/\pi \hbar^2 a_m$ corresponding to the Thomas Fermi radii
\begin{equation}
R_i^\textrm{BEC}=\tilde{a}_\textrm{ho}^{BEC} \left(\frac{15 N a_m}{2 \tilde{a}_\textrm{ho}^{BEC}}\right)^{1/5}
\frac{\tilde \omega_i}{\tilde \omega_\textrm{ho} },
\end{equation}
where $\tilde \omega_\textrm{ho}=(\tilde \omega_x \tilde \omega_y \tilde \omega_z)^{1/3}$ and $\tilde{a}_\textrm{ho}^{BEC}=\hbar/\sqrt{2m \tilde \omega_\textrm{ho}}$.
Notice that $\tilde{a}_\textrm{ho}^{BEC}$   differs from the oscillator length $\tilde a_\textrm{ho}$ defined above for the resonant case.

We now take advantage of the fact that the trapping potentials of the two different species can be tuned
{\sl separately} to suggest an experiment pointing out in a direct way the attractive role of the interactions in the presence of a Feshbach resonance. After generating the equilibrium configuration discussed above  we  switch off the confining potential  of a single species, say $V_\textrm{ho}^\downarrow =0$, 
corresponding  to a change of the trapping frequencies (\ref{weff}) into the new values
\begin{equation}\label{asi}
\tilde \omega_{i \textrm{new}} = \sqrt{\frac{m_\uparrow}{m_\uparrow +m_\downarrow}} \omega_{i \uparrow}.
\end{equation}
The potential can be switched off either adiabatically, bringing the system into a new equilibrium configuration, or suddenly, giving rise to the excitation of collective oscillations (see discussion in the second part of this work).

In the absence of interactions, the $\downarrow$-atoms would fly away leaving  
an ideal gas of  $\uparrow$-fermions trapped in the harmonic potential $V_\textrm{ho}^\uparrow$. This will be also the case in the deep BCS superfluid regime where interactions are too weak to keep the $\uparrow$-fermions confined.
In the other (more robust)  superfluid regimes, however, the released atoms do not necessarily escape to infinity but can remain trapped due to the attractive interaction with the other species. 
This statement is  obvious in the BEC regime, where each $\downarrow$-fermion forms a  bound molecule
with a corresponding $\uparrow$-particle. At unitarity, however, the two-body binding energy $E_b=-\hbar^2/2m_ra^2$ 
vanishes meaning that no molecules can exist in vacuum. In this case the  trapping of the $\downarrow$ 
component is a pure many-body effect reflecting the {\sl attractive} nature of the interatomic force.

It is not difficult to derive explicit conditions for the energetic stability of the new  superfluid configuration. The stability is ensured if the energy $E_S$ of the  configuration where the two components remain trapped and fully overlapped 
is smaller than the energy $E_N$ of the normal state where the gas is phase separated and only the $\uparrow$-atoms
are trapped. At unitarity 
the energy of the trapped superfluid state is given by 
\begin{equation}
\label{ene}
E_S=  \hbar \tilde \omega_\textrm{ho}     \frac{(3N)^{4/3}}{8}\sqrt{\frac{(m_\uparrow +m_\downarrow)}{m_r}}\sqrt{ 1+\beta}
\end{equation}
and, in the absence of trapping for the $_\downarrow$-atoms, one has $\tilde \omega_{ho}=\sqrt{m_\uparrow/(m_\uparrow+m_\downarrow)}(\omega_{x\uparrow} \omega_{y\uparrow}\omega_{z\uparrow})^{1/3}$.
Since the energy of a trapped gas of non-interacting $\uparrow$ fermions is given by $E_N=\hbar(\omega_{x\uparrow} \omega_{y\uparrow}\omega_{z\uparrow})^{1/3}(3N)^{4/3}/8$, we find that the 
superfluid gas, where both components remain trapped,  is {\sl energetically} stable if the condition 
\begin{equation}\label{cond}
 (1+\beta) \frac{(m_\uparrow +m_\downarrow)}{m_\downarrow}<1
\end{equation}
is satisfied.  Taking into account that $\beta$ barely depends on the mass ratio  \cite{stefano} and hence remains  close to the equal mass value $\beta=-0.58$, the above condition is always satisfied if 
$m_\uparrow \le m_\downarrow$, showing that the superfluid remains energetically stable
if we release the potential of the heavy species. 
Conversely, if the condition (\ref{cond}) is violated, the superfluid configuration corresponds to a metastable state which is {\sl energetically} unstable toward phase separation. 
It is also interesting to compare the Thomas Fermi radii $R_{iS}$ and $R_{iN}$ 
of the trapped cloud in the (new) superfluid and in the separated normal phase, respectively. A simple calculation yields
$R_{iS}/R_{iN}=(1+\beta)^{1/4} ((m_\uparrow +m_\downarrow)/m_\downarrow)^{1/4}$, showing that the superfluid phase corresponds
to the configuration with smaller radii if and only if Eq.(\ref{cond}) is satisfied.

Let us now discuss the macroscopic dynamic behavior of the superfluid. This is 
is obtained by deriving  
the hydrodynamic equations of motion  in terms of the atom density $n(\mathbf r,t)$ and the velocity field $\mathbf{v}(\mathbf r,t)$.
The Lagrangian $L$ of the system, in the local density approximation (LDA), is given by
\begin{equation}
L=\int d\mathbf r 
\Big[e(n)+(V_\textrm{ho}^\uparrow+ V_\textrm{ho}^\downarrow)\frac{n}{2}
+\frac{n}{2}\frac{\partial \phi }{\partial t}+\frac{1}{2} (m_\uparrow +m_\downarrow) \mathbf v^2 \frac{n}{2}
\Big]
\label{lag}  
\end{equation}
where $\phi$ is the phase of the order parameter $\langle
\Psi _{\uparrow }({\mathbf r},t)\Psi _{\downarrow }({\mathbf r},t)\rangle \equiv 
|\langle \Psi _{\uparrow }({\mathbf r})\Psi _{\downarrow }({\mathbf r})\rangle|
e^{i\phi ({\bf r},t)}$, $\Psi _\sigma(\mathbf r,t)$ being the fermionic field
operators.
The superfluid velocity $\bf v$ and the phase $\phi$ entering the Lagrangian (\ref{lag}) are related by the most important condition
\begin{equation}\label{velo}
\mathbf{v}=\frac{\hbar }{m_\uparrow+m_\downarrow}\nabla \phi,
\end{equation}
where $m_\uparrow+m_\downarrow$ is the mass of the pair, 
which can be derived microscopically by noticing  the state moving with velocity $\mathbf{v}$ is obtained from the steady state
by applying the gauge transformation $\Psi_{\sigma}(\mathbf r) \rightarrow \Psi_{\sigma}(\mathbf r)
e^{i m_\sigma \mathbf{v \cdot r}/\hbar }$  to the two Fermi field operators \cite{orso}. 

Taking Eq.(\ref{velo}) into account, the equations of motion of Lagrangian (\ref{lag})
yield the hydrodynamic equations
\begin{eqnarray}
&&\frac{\partial n}{\partial t}+\nabla(n\mathbf v)=0,\label{mass}\\
&& m \frac{\partial \mathbf{v} }{\partial t} + \nabla\left (\mu(n)+\tilde V_{ho}(\mathbf r)
+\frac{1}{2}m v^2\right)=0, \label{current} \; . 
\end{eqnarray}
Equations (\ref{mass}) and (\ref{current}) apply  to weakly as well as to strongly interacting superfluids and permit to calculate the macroscopic
dynamics of the system (expansion and collective oscillations) once the equation of state is known. 

In uniform matter ($\tilde V_\textrm{ho}=0$) the linearized solutions are phonons 
with sound velocity fixed by $c^2=(n/m) \partial \mu/ \partial n$.
At unitarity, the sound velocity is given by $c=\hbar k_F \sqrt{1/(3 m_\uparrow m_\downarrow)}\sqrt{1+\beta}$,
where $k_F=(3 \pi^2 n)^{1/3}$. In the BCS limit, the same formula holds with $\beta=0$ \cite{cina}.


For trapped configurations, the frequencies of the collective oscillations are obtained \cite{sandro96} by linearizing 
the hydrodynamic equations around the equilibrium distribution $n_0(\mathbf r)$   and depend
on the applied external potentials only through the effective oscillator frequencies (\ref{weff}). They  also depend on the equation of state and in particular on its density dependence \cite{sandro04}.
A special case is the center of mass oscillations of the cloud whose frequency, in the superfluid phase, is independent of the equation of state and given, for each direction,  by $\tilde \omega_i$. 
We emphasize that in general  $\tilde \omega_i$ differ from either $\omega_{i \uparrow}$ and $\omega_{i \downarrow}$, pointing out that in the superfluid phase the two components always oscillate in phase  as a consequence of the pairing mechanism. 

Equations (\ref{mass}) and (\ref{current}) remain valid even if the effective trapping frequencies 
depend explicitly on time, i.e. if $\tilde \omega_i=\tilde \omega_i(t)$. Remarkably,
if the chemical potential has the power law dependence $\mu \propto n^{\gamma}$ on the density
and the confining potential is harmonic, the hydrodynamic equations admit a class of exact  solutions given by $\mathbf v(t,\mathbf r)=\boldsymbol{ \alpha} \cdot \mathbf  r$ and $n(t,\mathbf r)=n_0(x/b_x,y/b_y,z/b_z)/(b_x b_y b_z)$, where $\alpha_i(t),\beta_i(t)$ are time-dependent parameters. From the scaling form of the density, we see
that the parameters $\beta_i(t)$ are related to the Thomas Fermi radii $R_i(t)$ of the evolving cloud 
according to  $R_i(t)=R_{i}(0)b_i(t)$.
Inserting the scaling ansatz in Eq.(\ref{mass})  yields the relationships  $\alpha_i=\dot{b}_i/b_i$ and 
one then obtains a set of coupled ordinary differential equations \cite{castin}
\begin{equation}\label{bi}
\ddot{b}_i+ \tilde \omega_i^2(t) b_i=\frac{\tilde \omega_i^2(0)}{b_i}\frac{1}{(b_x b_y b_z)^\gamma}.
\end{equation}
Equations (\ref{bi})  apply both to resonance, where $\gamma=2/3$, and to the BEC regime where $\gamma=1$.

If one suddenly switches off the trapping potential for the $_\downarrow$-component, corresponding to suddenly setting the effective frequencies to the new values (\ref{asi}), the superfluid gas will start oscillating around the new equilibrium configuration.  The dynamics of the Thomas Fermi radii
of the cloud can be calculated from Eq.(\ref{bi}) by substituting $\tilde \omega_i(t>0)= 
\tilde \omega_{i \textrm{new}} $ and employing the  initial conditions $b_i(0)=1$ and $\dot b_i(0)=0$. 
\begin{figure}[tb]
\begin{center}
\includegraphics[height=\linewidth,angle=270,clip]{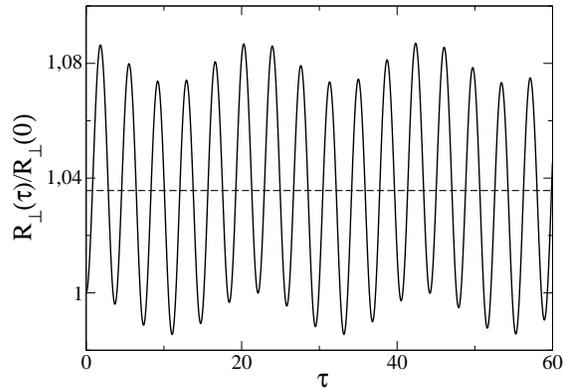}
\caption{Time evolution of the size of the cloud along the radial direction  after switching off suddenly the trapping potential for $^{40}$K fermions. Here $\tau=\tilde \omega_\perp (0) t$. The dashed line corresponds to the new equilibrium value (see text).}
\label{fig1}
\end{center}
\end{figure}

For simplicity, let us assume that the two trapping potentials  are axis-symmetric ($\omega_{x\sigma}=\omega_{y\sigma}\equiv\omega_{\perp \sigma}$). We further assume that the two components have initially equal oscillator lengths (and hence, in the degenerate limit, equal density profiles also in the absence of interaction), 
corresponding to $\tilde \omega_\perp^2(0)=\omega_{\perp \uparrow} \omega_{\perp \downarrow}$ and
$\tilde \omega_z^2(0)=\omega_{z \uparrow} \omega_{z \downarrow}$.
By introducing the dimensionless time $\tau=\tilde \omega_\perp(0) t$ and setting
$b_x=b_y=b_\perp$, Eqs (\ref{bi}) take the form
\begin{eqnarray}
 \frac{\partial^2 b_\perp}{\partial \tau^2}+  \eta b_\perp=\frac{1}{b_\perp} \frac{1}{(b_\perp^2 b_z)^{2/3}},\label{bx}\\
\frac{\partial^2 b_z}{\partial \tau^2}+ \eta \lambda^2 b_z=\frac{\lambda^2}{b_z} \frac{1}{(b_\perp^2 b_z)^{2/3}},\label{bz}
\end{eqnarray}
where $\eta=m_\downarrow/(m_\uparrow +m_\downarrow)$ and $\lambda=\tilde \omega_z (0)/\tilde \omega_\perp(0)$ is the aspect ratio of the  trapping potential (\ref{vtilde}). 
\begin{figure}[tb]
\begin{center}
\includegraphics[height=\linewidth,angle=270,clip]{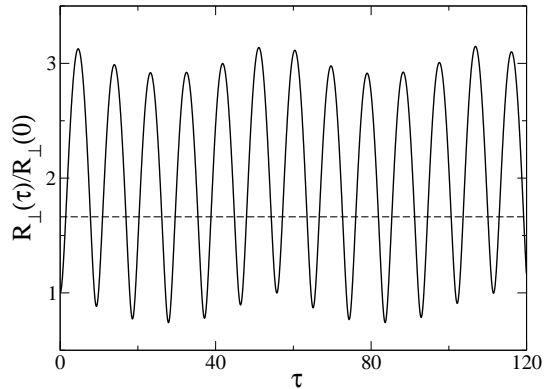}
\caption{Time evolution of the size of the cloud along the radial direction  after switching off suddenly the trapping potential for $^6$Li fermions. Here $\tau=\tilde \omega_\perp (0) t$. The dashed line corresponds to the new equilibrium value (see text).}
\label{fig2}
\end{center}
\end{figure}

We have solved Equations (\ref{bx}) and (\ref{bz}) for a mixture of $^{40}$K and $^6$Li fermions in an elongated trap with $\lambda=0.2$. In Fig.\ref{fig1} we plot the calculated time evolution of  the radius of the cloud in the radial direction, after we suddenly switch off  the trapping potential of $^{40}$K fermions, corresponding to $\eta=0.869$. The static solution $b_\perp=b_z=1/\eta^{1/4}$
of these equations and the corresponding  
 Thomas-Fermi radii  $R_{i \textrm{new}}=R_i(0)/\eta^{1/4}=1.036R_i(0)$ characterize the new equilibrium conditions after releasing the $\downarrow$ trapping potential.
Since $\eta$ is close to $1$, the initial configuration is close to equilibrium and 
the resulting oscillation is a linear superposition of the two breathing modes, in the radial and longitudinal directions. 
For elongated cloud, corresponding to $\lambda \ll 1$, these are given by  $\omega_\textrm{rad}=\sqrt{10/3}\eta^{1/2} \tilde \omega_\perp(0)$ and 
$\omega_\textrm{axial}=\sqrt{12/5}\eta^{1/2} \lambda \tilde \omega_{z}(0)$ \cite{rmp}.


In Fig.\ref{fig2} we instead plot the time dependence of  the transverse  radii of the cloud,  after we suddenly switch off  the trapping potential of the $^6$Li fermions, corresponding to $\eta=0.131$ and $R_{i \textrm{new}}=1.66 R_i(0)$.
In this case, the initial configuration is far from equilibrium and non linear effects play an important role.
In particular we see that the breathing of the cloud around the equilibrium configuration 
is no longer symmetric.
We emphasize that in this second case the superfluid configuration is energetically unstable. A major question in this case is to understand the decay mechanisms  and the role played by the sudden excitation of the collective modes.

In conclusions, we have derived the equilibrium conditions and hydrodynamic equations of a superfluid Fermi gas with unequal masses and
investigated the behavior of a   gas  at unitarity after the release of the trapping 
potential of a {\sl single} component. We have shown that, under appropriate conditions, the
 superfluid phase is energetically stable against phase separation of the two components. As a result, the released fermions
remain confined in the trap due to the pairing with the other component, pointing out in a direct and remarkable way the attractive nature of the interatomic forces near a  Feshbach resonance.

This work is supported by the Marie Curie program under contract EDUG-038970 and by
the Ministero dell'Istruzione, dell'Universit\`a e della Ricerca (M.I.U.R.).

\end{document}